\documentstyle[aps,multicol,pre,eqsecnum,epsf]{revtex}

\newcommand{\equ}[1]{(\protect\ref{#1})}
\newcommand{\Prl}[3]{Phys. Rev. Lett. {\bf #1}, #2 (#3).}
\newcommand{\Jpa}[3]{J. Phys. A {\bf #1}, #2 (#3).}
\newcommand{\Generic}[3]{{\bf #1}, #2 (#3).}
\newcommand{\ep}{\varepsilon} 
\newcommand{\al}{\alpha}
\newcommand{\be}{\beta}
\newcommand{\cT}{{\cal T}(N)}

\newcommand{\cR}{{\cal R}(N, m, \sigma)}
\newcommand{\cS}{{\cal S}(N, L)}
\newcommand{\cD}{{\cal D}(N, t_0)}
\newcommand{\cCT}{{\cal C_T}(N)}

\newcommand{\reals}{{\rm I \mkern-2.5mu \nonscript\mkern-.5mu R}}
\newcommand{\Raya}{\noindent\makebox[3.4truein]{\hrulefill}}
\newcommand{\Mycap}[1]{\protect\noindent\protect\parbox{8.6cm}{\small #1}}

\begin{document}

\draft

\title{Multifractal properties of power-law time sequences; application to 
ricepiles}

\author{Romualdo Pastor-Satorras\cite{email}}

\address{Department of Earth, Atmospheric, and Planetary Sciences\\
Massachusetts Institute of Technology\\
77 Massachusetts Avenue, Cambridge, Massachusetts 02139}

\maketitle

\begin{abstract} 
We study the properties of time sequences extracted
from a self-organized critical system, within the framework of the
mathematical multifractal analysis. To this end, we propose a
fixed-mass algorithm, well suited to deal with highly inhomogeneous one
dimensional multifractal measures.
We find that the fixed mass ({\em dual\/})
spectrum of 
generalized dimensions depends on both the system size $L$
and the length $N$ of the sequence considered, being however
stable  when 
these two parameters are kept fixed. A finite-size scaling relation is
proposed, allowing us to define a renormalized spectrum, independent
of size effects.
We interpret our results as an
evidence of extremely long-range correlations induced in the sequence
by the criticality of the system
\end{abstract}


\begin{multicols}{2}

\section{Introduction}

Self-organized criticality (SOC) has been the subject of a great deal
of interest, since its introduction by Bak, Tang, and Wiesenfeld
\cite{bak87}. The main feature of SOC systems is that they
evolve, driven by means of an external force, into a critical state
characterized by the absence of any characteristic time or length
scale. The resulting extremely long range correlations  show up
through the  peculiar ``$1/f$'' power spectrum and the geometrical
fractal structure.
SOC behavior has been observed in many cellular automata models of
sandpiles \cite{bak87},
invasion percolation \cite{roux89}, biological evolution \cite{bak93},
depinning in random media \cite{sneppen92}, and also in some natural
systems, such as earthquakes \cite{bak89}. Even though  the first
cellular automaton displaying SOC was conceived to represent the
dynamics of a sandpile \cite{bak87}, the experimental evidence 
indicates that this is not actually the case: Real sandpiles are not in a
self-organized critical state \cite{held90,rosendahl94,feder95}.
Recently, however, Frette {\em et al.} \cite{frette96} reported 
SOC behavior in a real granular system, a one dimensional
ricepile. For grains of rice with a considerable 
aspect ratio, the pile behaves critically, this fact being accounted for
by the increased friction, which is able to counterbalance the
inertia effects predominant in real sandpiles. 

In a subsequent paper,
Christensen  {\em et al.}  \cite{christensen96} analyzed the transport
properties of individual grains inside a stationary ricepile. They
measured the
{\em transit time} of individually colored grains of rice (tracers),
defined as the time 
necessary for a grain to escape from the pile. Christensen  {\em
et al.} found that the distribution of transit times follows a
truncated power law form, and that the average transport velocity of
the grains diminishes as the system size increases. A cellular
automaton model of a ricepile was  proposed (the so-called Oslo model)
\cite{christensen96,corral97}, reproducing the phenomenological
behavior of the actual experiments. Bogu\~{n}\'a and
Corral~\cite{bogunya97} have also suggested a theoretical scenario for
the Oslo ricepile, based on a continuous time random walk model.

The main results of the Oslo experiments and simulations  can be
expressed through a single 
function,  the probability distribution of transit times $P(T, L)dT$, 
which is defined as the probability of a  given tracer spending 
a time between $T$ and $T+dT$ inside   
a pile of linear size $L$. It was found that
\begin{eqnarray}
&&P(T, L) \sim L^{-\nu}, \quad T < L^\nu,\nonumber \\
&&\label{properties} \\
&&P(T, L) \sim T^{-\chi}, \quad T > L^\nu, \nonumber
\end{eqnarray}
where $\nu$ and $\chi$ are certain characteristic exponents. The
experiments provided the values $\nu=1.50\pm0.20$ and
$\chi=2.40\pm0.20$, whereas the cellular automaton model rendered the
exponents $\nu=1.30\pm0.10$ and $\chi=2.22\pm0.10$
\cite{christensen96,corral97}. 
This numerical evidence  can be summarized in the finite-size scaling ansatz
\begin{equation}
P(T, L) = L^{-\be} f\left( \frac{T}{L^\nu} \right),
\label{ansatzzz}
\end{equation}
with
\begin{equation}
f(x)=\left\{ \parbox{1.2in}{%
		\makebox[0.5in][l]{const.} for $ x < 1$ \\
		\makebox[0.5in][l]{$ x^{-\chi}$} for $ x > 1$} \right. .
\end{equation}
Given that $\chi>2$, and provided that the probability distribution is
normalized, we have that $\be=\nu$ and the average value of $T$ is
finite, $\left<  T \right> \sim L^\nu$ (see
Appendix).  The fact that $\chi<3$
implies,  however, that the second moment of the distribution is
infinite, $\left< T^2 \right> = \infty$.  

The finite-size scaling~\equ{ansatzzz}  compacts the experimental data into
a useful relationship, which on its turn allows one to extract valuable
conclusions about the system. However,  it is actually quite obvious
that it is possible to extract more information about the ricepile
from the {\em sequence} of transit times, apart from its distribution
function. 
In order to gain a different insight into the problem, we propose to
consider the output of the experiment from a different point of view. 
Let us
define the set  $\cS$ as follows:
Throw a tracer grain in a stationary pile \cite{stability} 
of linear  size $L$ and measure the time elapsed until it emerges outside
within any  avalanche.  Performing the same measurement for $N$
different grains, consecutively  thrown in the pile, we can
construct  the sequence $\cS=\left\{ T_n  \right\}_{n=1,\ldots,N}$,
where $T_n$ is the time, measured in units of added grains (the slow
time scale~\cite{christensen96}),
spent inside the pile  by   the $n$-th grain,
in the sequence of $N$ {\em consecutive} throws. The set $\cS$ can be
interpreted as a discrete time sequence, 
assigning   to the instant $n=1,\ldots,N$ the value $T_n$. 
In the Fig.~1 we have represented such a sequence, for the transit times
recorded in the cellular automaton model of ricepile described in
Ref.~\cite{christensen96}. The system size is
$L=100$. Fig.~1(a) 
shows a record of 90000 transit times, whereas Fig.~1(b) depicts the 5000
points closer to the center of the previous graph. These plots show
rather conclusively that not only the {\em distribution} of transit times is
scale-free, but also that their   {\em sequence}
is in some sense  self-similar.


\begin{figure}[t]
\epsfxsize=8truecm
\centerline{
\epsfbox[60 45 540 740]{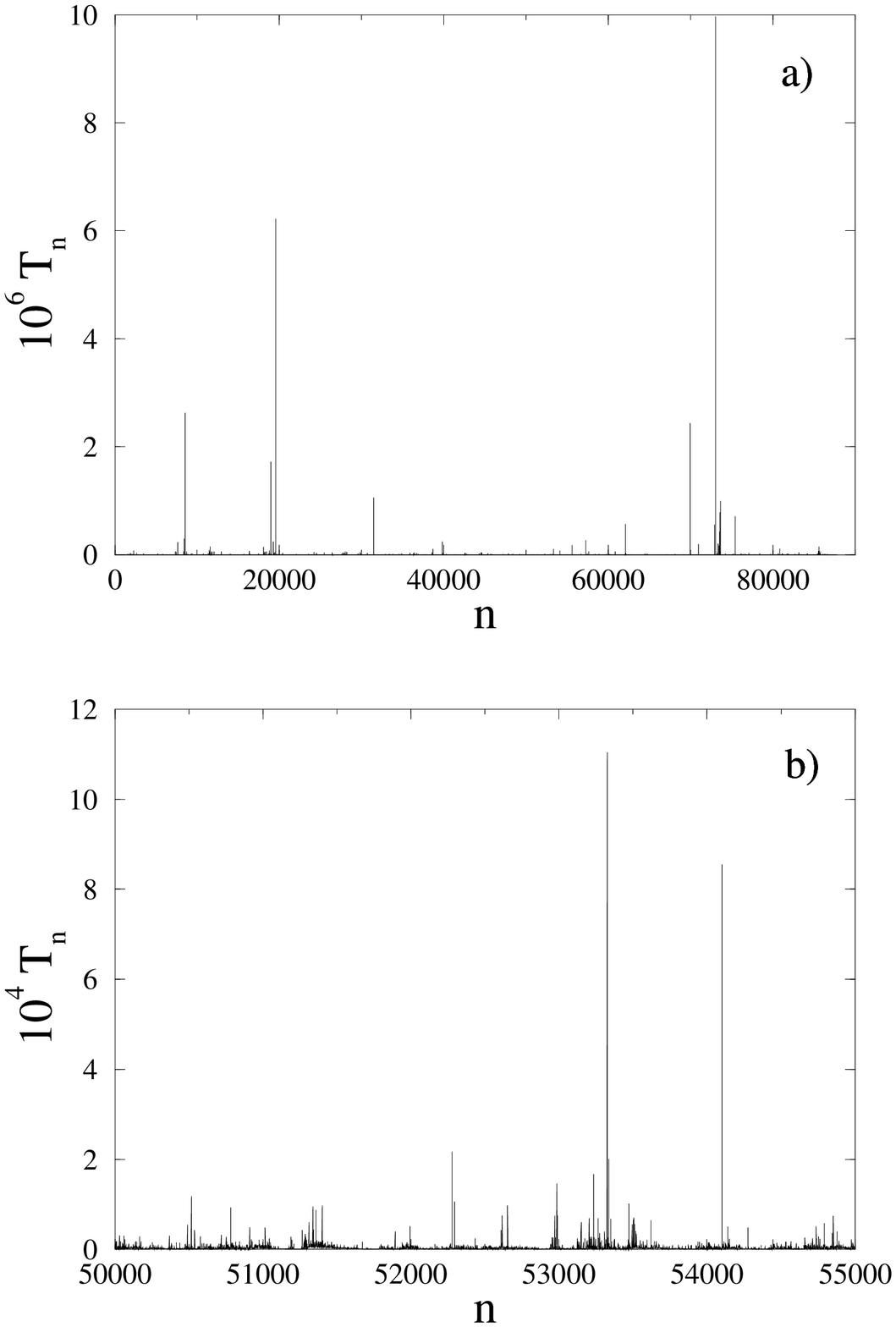}
}
\Mycap{Fig. 1: a) Sequence of transit times for 90000 tracer grains
in a computer simulation of the Oslo ricepile; system size $L=100$. b)
Zoom close to the central section of the previous picture.}
\end{figure}

In this paper we will  extract  more information from the
Oslo ricepile model studying the sequence of transit times
$\cS$. The method we have employed is that of {\em multifractal analysis}
(which, on the other hand, is not new in the field of SOC 
\cite{kadanoff89,olami92}). To this end, we have developed an
algorithm particularly well suited to deal with one dimensional
measures, like the ones under consideration. When computing the
multifractal spectrum of the sequence $\cS$, we observe that it shows
considerable size effects: The spectrum of generalized dimensions
$D(q)$ (to be defined later on) depends on the system size $L$ and,
even worse, on the sequence length $N$. This fact seems to doom any
effort to describe a single well-defined spectrum. However, by
analyzing $D(q)$ in the limit $q\to\infty$, we observe a power
law dependence on $N$ and $L$. Extending the scaling to the whole
range of $q$ allows us to define a ``renormalized'' spectrum, truly
independent of size effects. We interpret our results as an effect of
the extremely long ranged correlations  present in the sequence,
correlations induced by the criticality of the ricepile.

We have  organized  this manuscript as follows: In Sec.~II we
review  the multifractal analysis of  general
mathematical measures,
stressing the difference between fixed-size and fixed-mass
formalisms. In Sec.~III we develop in particular the formalism needed
to deal with 
a discrete time sequence. Sec.~IV analyses different synthetic uncorrelated
random time sequences. First we check the accuracy of the algorithm
against sequences of known spectra. Then we study a power-law
distributed random signal, mimicking the real transit time sequences.
Section~V
deals with our final goal, actual sequences of transit times from numerical
simulations of the Oslo model. Finally, our conclusions are discussed
in Sec.~VI. 

\section{Multifractal analysis: fixed-size vs. fixed-mass formalism} 

Loosely speaking, we call {\em multifractals}
\cite{mandelbrot74,benzi84,halsey86} the
mathematical sets  which can be 
decomposed into an infinite set of interwoven subfractals, labelled with
an index $\al$, each of them
characterized by a different fractal dimension $f$. The collection of these
dimensions form the so-called {\em multifractal spectrum} $f(\al)$
\cite{falconer94}. Strictly speaking, however, it is only possible to
assign mathematically meaningful multifractal properties to a {\em
measure} (mathematical or physical) defined over a given support
\cite{falconer94}. 
A multifractal measure is completely specified either by its
multifractal spectrum $f(\al)$ or by its spectrum of generalized
dimensions $D(q)$. 

In this section we review the main mathematical
definitions and properties of multifractal analysis.

\subsection{General definitions}

Following Ref.~\cite{falconer94} (see also \cite{mach95}), consider a
normalized  measure $\mu$ defined on a support $K \subset \reals^d$,
$\mu(K)=1$. Let $\Delta$ be an arbitrary partition of $K$ in
non-intersecting elements $\Delta_i$, that is,
\begin{equation}
K \subseteq \bigcup_i \Delta_i \qquad \mbox{\rm and}  \qquad \Delta_i
\bigcap \Delta_j = \emptyset, \, i \neq j,
\end{equation}
and
let $p_i$ and $\ep_i$, $i=1,\ldots,N$ be the  variables that
represent the weight factor  and the size factor corresponding to
the element $\Delta_i$, respectively. We define the function 
\begin{equation}
\Phi_\Delta(q, \tau) = \left< \sum_{i=1}^N p_i^q \ep_i^{-\tau} \right>,
\end{equation}
where $q$ and $\tau$ are any real numbers.  The sum runs over all the $N$
disconnected 
parts in which we  decompose the support of the measure, and the
brackets stand for an average over different realizations of the
measure. For any measure, either deterministic or experimental (non
deterministic), we will assume that, for fine enough partitions,  the
function $\Phi_\Delta(q, \tau)$ 
collapses onto a single constant value, that is, 
\begin{equation}
\left< \sum_{i=1}^N p_i^q \ep_i^{-\tau} \right> = \mbox{\rm const.} 
\label{low-level-partition}
\end{equation}
Expression~\equ{low-level-partition} is an implicit equation, allowing
one to determine $\tau(q)$ for a given $q$, or conversely, $q(\tau)$
for a given $\tau$.
If we assume a partition $\Delta$ in which  $\ep_i = \ep = \mbox{\rm
const.}$, then the size factor $\ep$ 
can be 
factorized from the former expression, yielding  
\begin{equation}
\left< \sum_{i=1}^{N(\ep)} p_i^q \right> \sim \ep^{\tau},
\label{fixed-size}
\end{equation}
where $N(\ep)$ is the number of parts of size $\ep$, containing a
certain measure $p_i$ different from zero. From this last expression we
can compute the function  $\tau(q)$ and the {\em generalized
dimensions} $D(q)$ \cite{grassberger83,hentschel83}, defined by 
$D(q)=\tau(q)/(q-1)$.  The  $f(\al)$ spectrum is given by the
Legendre transformation $f(\al)=\min_q\{q\al -(q-1)D(q)\}$
\cite{falconer94,cawley92}. 
This approach corresponds to the
so-called {\em fixed-size multifractal formalism} (FSF). 
  
On the other hand, we can select a partition $\Delta$ in which $p_i = p =
\mbox{\rm const.}$, which 
yields to
\begin{equation}
\left< \sum_{i=1}^{N(p)} \ep_i^{-\tau} \right> \sim p^{-q},
\label{fixed-mass}
\end{equation}
where $N(p)$ is the number of parts of measure $p$, with a certain
size  $\ep_i$ different from zero. From this expression we can
calculate the function  $q(\tau)$,  
and then, inverting it, compute the spectrum
$D(q)$. This second  
approach corresponds to the so-called {\em fixed-mass multifractal
formalism}  (FMF). 

Both FSF and FMF are completely equivalent. In order 
to stress this  correspondence, we 
define the new parameters $q^*\equiv-\tau$ and $\tau^*\equiv-q$ and
substitute into Eq.~\equ{fixed-mass}.  Now both
equations  
\equ{fixed-size} and \equ{fixed-mass} read the same,
 the only
difference being the change of role of $p_i$ and 
$\ep_i$.  The equivalence between both formalisms is explicitly illustrated
by the identities  
\begin{eqnarray}
q^* & = & -(q-1) D(q),\nonumber \\
&& \label{translation}  \\
D^*(q^*) & = & \frac{q}{1 + (q-1) D(q)}, \nonumber
\end{eqnarray}
with $D^*(q^*)=\tau^*(q^*)/(q^*-1)$. 

\subsection{Box-counting algorithms}

The most common operative  numerical implementations of 
multifractal analysis are  the 
so-called {\em fixed-size  box-counting
algorithms}~\cite{halsey86}. For a given measure $\mu$ with support $K
\subset \reals^d$, they consider the {\em partition sum}
\begin{equation}
Z_\ep(q) = \sum_{\mu(B)\neq0} \Big( \mu(B) \Big)^q,
\end{equation}
$q\in\reals$, where the sum runs over all the different non-empty
boxes $B$ of 
a given side  $\ep$ in a $\ep$-grid  covering the support $K$, that
is
\begin{equation}
B=\prod_{k=1}^{d} \; \left]l_k \ep, (l_k + 1) \ep\right]\;,
\label{equ-B}
\end{equation}
$l_k$ being integer numbers.
The generalized  fractal dimensions of the measure are defined by
the limit
\begin{equation}
D(q) = \frac{1}{q-1} \lim_{\ep\to0}\frac{\log Z_\ep(q)}{\log \ep} 
\label{mathemat}
\end{equation}
and numerically estimated through a linear regression of 
\begin{equation}
\frac{1}{q-1} \log Z_\ep(q)
\end{equation}
against $\log \ep$.

Within this formalism, the mathematical definition~\equ{mathemat} is
strictly valid for  positive  $q$ \cite{riedi95}.  Numerical
estimates work well for $q>1$ in  $d>2$, and render usually incorrect
results for $q<0$ \cite{grassberger88,greenside82,pastor96}.  This
fact is  
obviously due to the presence of boxes $B$ with an unnaturally small
measure, which contribute to the function $Z$ with diverging terms. 
In those cases, one is forced to apply different  
prescriptions  \cite{pastor96,alber97}.

The box-counting version of the fixed-mass formalism is in general
harder to implement in $d>1$ spatial dimensions. The difficulties reside
in the proper selection of boxes  with a given
fixed measure. 
(For an application in $d=2$ see Ref.~\cite{mach95}.) 
From a numerical point of view,
it is well known that the FMF is a good estimator of generalized dimensions
for $q<0$ (that is, $q^*>0$, see Eq.~\equ{translation}) and bad for  $q>0$  
($q^*<0$). The explanation of this behavior is
related to the space distribution  
of the measure. The FSF operates well in the dense regions of the
support,   whereas the FMF is specially appropriate to deal with its
sparse regions. 

As we will see in the next section, however, a fixed-mass algorithm is
particularly simple to implement for one dimensional measures, such
as time sequences.

\section{Multifractal formalism for discrete time sequences}

Fractal geometry and multifractal analysis are well-known tools for
the study of complex time signals (see for instance
\cite{levy-vehel96,riedi96}  and references therein).  
In this section we will specialize the box-counting
multifractal analysis sketched above for the particular case of a
discrete one dimensional time sequence.

We define a general discrete time sequence $\cT$ as any set of $N$
positive real numbers, $\cT=\left\{ t_n \right\}_{n=1,\ldots,N}$,
$t_n\in\reals^+$. At this level we will not make any assumption about the
possible  correlations of the 
sequence. However, we will assume that it is the outcome of some physical
process in a stationary state, and that we can obtain sequences as long
as it might be required. 

\subsection{Fixed-size algorithm}
\label{naive}

In order to study the multifractal properties of a sequence $\cT$, we
must first provide a  meaningful physical measure on it. As a first
ansatz, we define the {\em naive  measure} $\mu$ on
the support $]0, N] \subset  \reals$ over which the sequence is
defined. This  measure   assigns to a given box in $]0,
N]$ a weight  proportional to the sum of the value $t_n$ of all the
points $n$ 
inside 
the box~\cite{riedi96}. Namely, if $B(x,\ep)$ is a ball with center in
$x$ and  diameter $\ep$, then
\begin{equation}
\mu(B(x,\ep)) = \frac{1}{Q(N)} \sum_{x-\frac{\ep}{2} < n \leq
x+\frac{\ep}{2}} t_n,
\end{equation}
where $Q(N) = \sum_{n=1}^N t_n$ is a normalization factor such that
$\mu(]0, N])=1$.  In order
to  compute the generalized dimensions $D(q)$ of $\mu$, consider a
partition  of  $]0, N]$ into boxes of diameter $r$, in a number $N/r$,
defined by
\begin{equation}
B_{k,r} = \;
](k-1) r, k r], \qquad k=1,\ldots,N/r. 
\end{equation}
The partition sum will then read 
\begin{equation}
Z_{r}(q) = \sum_{k=1}^{N/r} \Bigr(\mu(B_{k,r})\Bigl)^q =
\frac{1}{Q(N)^q} \sum_{k=1}^{N/r} \Bigr(\sum_{(k-1)r < n
\leq  kr} t_n \Bigl)^q.
\end{equation}
The generalized dimensions are defined through
\begin{equation}
D(q) = \frac{1}{q-1} \lim_{r/N\to0} \frac{\log \left\{\frac{1}{Q(N)^q}
\sum_{k=1}^{N/r}  \Bigr(\sum_{(k-1)r < n \leq kr} t_n  
\Bigl)^q \right\}}{\log\frac{r}{N}}.
\end{equation}
The role of $\ep$ is now played by the reduced diameter of the boxes
$r/N$.  Numerically, we will obtain an estimate of $D(q)$ as
the slope of a linear  regression of  
\begin{equation}
\frac{1}{q-1}  \log \sum_{k=1}^{N/r}
\Bigr(\sum_{(k-1)r < n \leq kr} t_n \Bigl)^q
\end{equation}
against $\log(r/N)$. Note that we have dropped the normalization factor
$Q(N)^q$,  since it does not depend on $r$ and therefore plays no
role in the regression. Moreover,  the elimination of this
factor results  in  general in a better performance of the numerical
algorithm, except for  those values of $q$ very close to $1$.

\subsection{Fixed-mass algorithm}
\label{non-naive}

In order to define a fixed-mass algorithm for a discrete sequence
$\cT$, we start by constructing an approximate Cantor set $\cCT$,
composed by a  
collection of $N$ discrete points on the interval $]0,1]$. 
We define the {\em dual measure} $\mu^*$ by associating a mass
distribution to this approximate Cantor set. The distribution
corresponds to just assigning a mass
unity to each 
one of its  points. Consider thus the sequence $\cT=\left\{ t_n
\right\}_{n=1,\ldots,N}$, with $Q(N)=\sum_{n=1}^N t_n$, 
and let us define the Cantor set $\cCT$ by
\begin{equation}
\cCT=\left\{ x_n \; | \; 0 < x_n \leq 1, n=1,\ldots,N \right\} 
\end{equation}
with
\begin{equation}
x_n=\frac{1}{Q(N)} \sum_{k=1}^n t_k. 
\end{equation}
We define a measure on  $\cCT$  through the density function
\begin{equation}
\rho_{\cal C}(x) = \frac{1}{N} \sum_{n=1}^N \delta(x - x_n),
\end{equation}
where $\delta$ is the usual Dirac delta function. The measure of a ball
of center $x$ and  diameter $\ep$, $B(x, \ep)=\;
]x-\frac{\ep}{2}, x+\frac{\ep}{2}]$ is given by the integral
\begin{equation}
\mu^*(B(x, \ep)) = \int_{x-\frac{\ep}{2}}^{x+\frac{\ep}{2}} \rho_C(x) dx
\end{equation}
and is equal to the number of points from $\cCT$  contained in
the  interval $B(x, \ep)$. It is easy to verify that the dual measure
$\mu^*$ has  holes of finite size: Consider a given $t_p$ and 
\begin{equation}
\bar{\ep} < \frac{t_p}{Q(N)},
\end{equation}
and define
\begin{equation}
\bar{x} = x_{p-1} + \frac{t_p}{2 Q(N)}.
\end{equation}
Then $\mu^*(B(\bar{x}, \bar{\ep})) = 0$. If $t_p$ is very large, then
it will  
correspond to a large hole in $\cCT$, with a  diameter 
$\frac{t_p}{Q(N)}$. This implies
that the fractal dimension of the support of $\mu^*$ would be in general
lesser than 
$1$. These regions of zero dual measure are related to the regions of
large naive measure. 

We define the FSF multifractal spectrum of $\mu^*$, $D^*(q^*)$,
through the partition function $Z^*_\ep(q^*)$, which on its turn is
defined onto the basis of a set of disjoint intervals covering
$]0,1]$:
\begin{equation}
B_{k,\ep} = \; ](k-1)\ep, k \ep], \qquad k=1,\ldots,1/\ep,
\end{equation}
that is,
\begin{equation}
Z^*_\ep(q^*) = \sum_{k=1}^{1/\ep} \mu^*\Bigl(B_{k,\ep}\Bigr)^{q^*}=
\frac{1}{N^{q^*}}  \sum_{k=1}^{1/\ep} \left[ \int_{(k-1)\ep}^{k\ep}
\rho_C (x) dx \right]^{q^*}.
\end{equation}
The generalized dimensions are mathematically defined by the limit
\begin{equation}
D^*(q^*)=\frac{1}{q^*-1} \lim_{\ep\to0} \frac{ \log Z^*_\ep(q^*)}{\log\ep}
\end{equation}
and numerically evaluated as the slope of a linear fit of 
\begin{equation}
\frac{1}{q^*-1} \log Z^*_\ep(q^*)
\end{equation}
against $\log \ep$. We will drop again the normalization factor
$N^{q^*}$.

From a mathematical point of view, this construction represents a
practical implementation of the notion of {\em inverse multifractal
measure} discussed by Mandelbrot and Riedi in Ref.~\cite{mandelbrot97}.
Let us show that $\mu^*$ indeed corresponds to the inverse
of the naive measure defined on the original
sequence. Consider a box 
$B_{k}$ of size $\ep_k$, which contains $n_{k}$ points from the  $\cCT$, and 
therefore has an associated dual measure 
\begin{equation}
\mu^*(B_{k})=\frac{n_{k}}{N}.
\end{equation}

Consider that those $n_{k}$ points are the consecutive points
$x_{l}, x_{l+1},\ldots, x_{l+n_{k}-1}$. Assuming that the extreme
points coincide with the extremes of the interval,
then we have $x_{l+n_{k}-1} - x_{l} \sim \ep_k$. If we
recover the former definition of $x_n$, then 

\end{multicols}

\widetext

\Raya\hfill

\begin{equation}
\ep_k  \sim  x_{l+n_{k}-1} - x_{l} = \frac{1}{Q(N)}
\sum_{s=1}^{l+n_{k}-1} t_s  - \frac{1}{Q(N)} \sum_{s=1}^{l} t_s
= \frac{1}{Q(N)} \sum_{s=l+1}^{l+n_{k}-1} t_s \simeq
\mu(\tilde{B}_k),
\end{equation}
\mbox{%
where $\tilde{B}_k$ is a certain box, associated with the naive
measure, with
diameter $\tilde{\ep}_k \sim n_{k} / N$. Then we have
}
\begin{equation}
\sum_k \mu^*(B_{k})^{q^*} \ep_k^{-\tau^*} = 
\sum_k \left( \frac{n_{k}}{N} \right)^{q^*} \ep_k^{-\tau^*} \sim
\sum_k \left( \tilde{\ep}_k \right)^{q^*} \mu(\tilde{B}_k)^{-\tau^*} =
\sum_k \mu(\tilde{B}_k)^q \tilde{\ep_k}^{-\tau}.
\end{equation}

\hfill\Raya

\begin{multicols}{2}

\noindent In the last equality we have identified $\tau = - q^*$ and
$q = - \tau^*$. 
We then see that computing the spectrum of $\mu^*$ by covering its
support with boxes of given size is the same as computing the spectrum
of $\mu$ by means of a covering of boxes of given mass. That is, one
measure is the inverse of the other, in the sense of \cite{mandelbrot97}.
Specializing to boxed of fixed size of mass, 
we can state  that computing the fixed-mass spectrum of
the naive measure $\mu$ on the sequence $\cT$ amounts to the
computation of the fixed-size spectrum of the dual measure $\mu^*$ on
the approximate Cantor set $\cCT$, and the other way around.

In the remaining of this paper we will focus mainly on the
spectrum of  the dual measure $\mu^*$ for the time sequence considered
({\em dual spectrum}), as opposite to the spectrum of the naive
measure ({\em naive spectrum}).  Therefore, in
order to alleviate notation we will
denote this particular dual spectrum and associated magnitudes without the
explicit star-superindex notation, unless otherwise stated.

\section{Numerical results for synthetic time sequences}

In this section we present our estimates for the multifractal
spectrum of some synthetic (computer generated)  time
sequences. First we check our algorithm with two measures of
known multifractal spectrum. Finally, we study the special case of a
random signal whose values are distributed according to a truncated
power law. 

The numerical procedure for computing estimates of dimensions
$D(q)$ implies the  {\em quenched average} of the partition sum over an
ensemble of statistically  independent realizations of the signal,
each one with the same length $N$. By quenched 
averages we refer to the mean value of the logarithm of the partition
sum,  $\left< \log Z_\ep(q)  \right>$. As it is well-known, this kind
of average is more stable and less subject to a particular sampling of  
scarce significance than the annealed average, which would consider 
the logarithm of the mean value of the partition sum, $\log \left< Z_\ep(q)
\right>$. In order to obtain
results comparable in a 
straight forward way for  any value of $q$, the linear regressions to
estimate $D(q)$ are always performed over the same scaling interval,
independently of the particular value of $q$  considered.

\subsection{Uniform random sequence}

Firstly, we analyze a uniform random sequence $\cR$, where the
different values  $t_n$ are uniform 
uncorrelated random variables with mean value $m$ and standard
deviation $\sigma$. For our numerical experiments we choose $m=100$
and $\sigma=10$. For a smooth signal like $\cR$ we expect to obtain a
flat multifractal spectrum, that is, generalized dimensions equal to
unity for both  naive and dual measures. This expectation is confirmed by
our computations, which yield generalized dimensions satisfying
$|D(q)-1|\leq0.001$ for $|q|\leq10$, and dimensions very close to $1$
for $10 < |q| \leq 40$

\subsection{Self-similar deterministic sequence}

We can construct a fully multifractal  sequence starting from any
self-similar  deterministic multifractal measure on
$\reals$~\cite{halsey86,falconer90}.  We considered a {\em fixed-size}
measure with contraction factor $r=1/2$ and probabilities $p_1=0.3$
and $p_2=0.7$ \cite{falconer90} and constructed a non-normalized
approximation of the 
measure composed by $1.1 \times 10^7$ points by means of the `Chaos
Game'~\cite{barnsley88}. The multifractal sequence was eventually
constructed by binning the sample points in $5 \times 10^4$ boxes covering the
interval $]0,1]$  over which the original measure was defined. The value
$t_n$ of the sequence is then given by the occupation number of the
$n$-th box. Fig~2(a) depicts such a sequence. Its self-similarity seems
obvious even to the naked eye. 

The analytical dual spectrum of the sequence is given,
as a function of the parameter $s\in\reals$, by the
expression~\cite{halsey86,falconer90} 
\begin{eqnarray}
q_s & = & -\frac{\log(p_1^s + p_2^s)}{\log(s)}, \nonumber \\
&&\label{analytic-spectrum}\\
D(q_s)  & = & \frac{s}{1 +\frac{\log(p_1^s + p_2^s)}{\log(s)} }. \nonumber
\end{eqnarray}
(Recall that $D(q)$ stands now the fixed-mass spectrum of the
original naive multifractal measure. The expression for its fixed-size 
spectrum, commonly found in the literature, is rather less complex.)
In our computations we averaged over
$10$ different approximations of the sequence. Linear regressions were
performed over an interval of $2.5$ decades. Error bars correspond to
statistical errors from the regression algorithm. 
In Fig~2(b) we have plotted our numerical estimates of the dual
spectrum for sequences of
length $N=10000$,  together with the analytic
spectrum~\equ{analytic-spectrum}. The figure shows
an excellent agreement between our estimates and the expected
analytic result, in the whole interval of values of $q$ considered,
both positive and negative. The accuracy of the fit can be slightly
improved by increasing the sequence length, but the estimates are
already 
quite stable and correct for the value of $N$ showed in the figure.
Computations performed for the naive spectrum yielded an equally good
agreement with the analytical result.


\begin{figure}[t]
\epsfxsize=8truecm
\centerline{
\epsfbox[20 80 500 710]{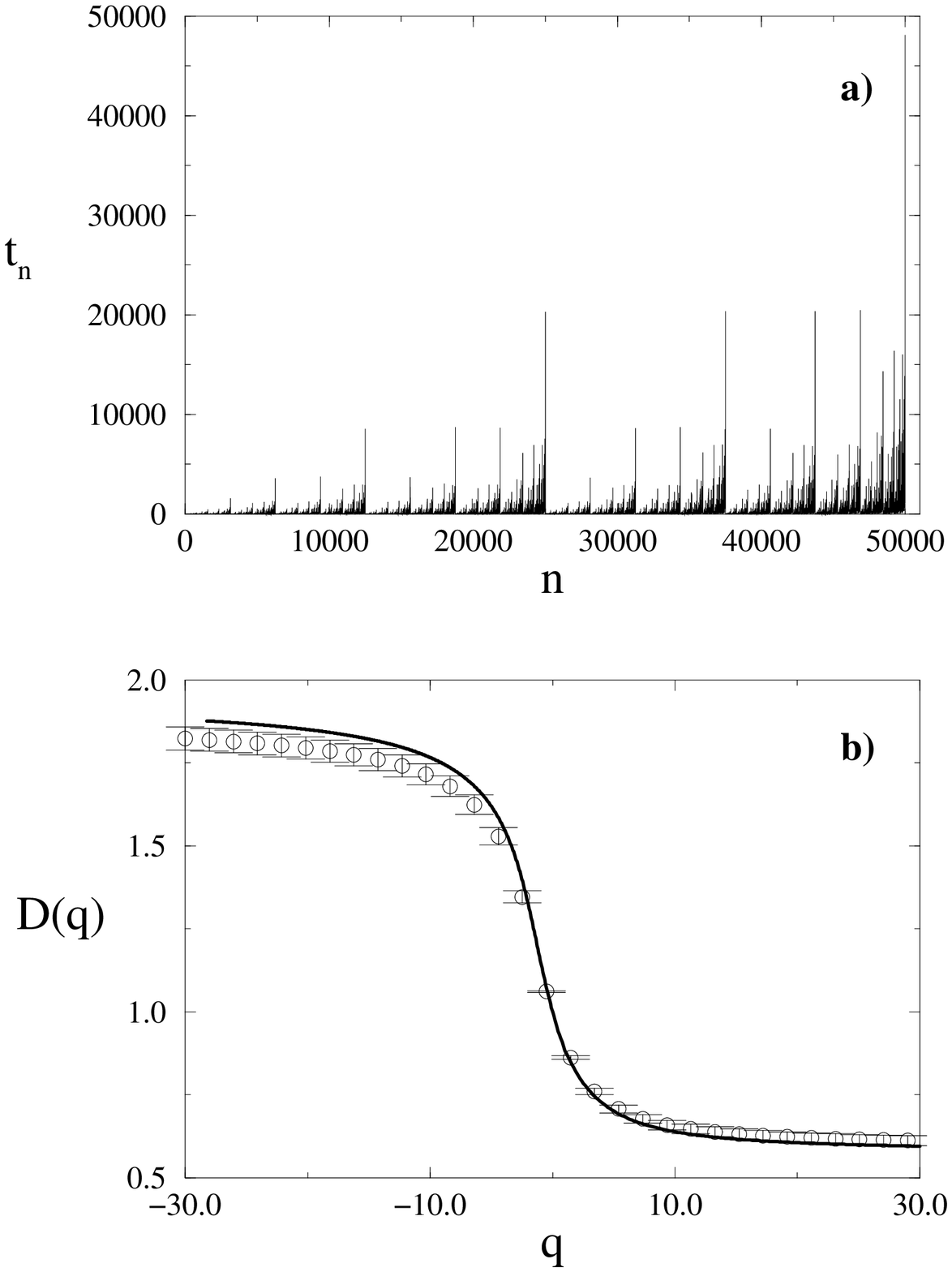}
}
\Mycap{Fig. 2:
a) Succession of $50000$ values from a deterministic multifractal time
sequence. Its parameters are $r=1/2$, $p_1=0.3$, and $p_2=0.7$ (see text).
b) Mathematical dual spectrum of the above sequence (full line); the
points represent our numerical estimates.}
\end{figure}

\subsection{Power-law random sequence}
\label{synthetic-power-law}

The sequence of transit times seems to be distributed according to a
truncated power-law of the form
\begin{equation}
\rho(t, t_0)=\left\{\parbox{1.3in}{%
			\makebox[0.5in][l]{const.} $t\in[0,t_0[$ \\
			\makebox[0.5in][l]{$a\; t^{-\chi}$} $t\in[t_0, \infty[$} 
			\right. .  
\label{akin-equ}	
\end{equation}
(see Eq.~\equ{properties}). In order to explore the applicability of
our algorithm to a power-law 
sequence, we have constructed and analyzed a synthetic random sequence $\cD =
\left\{ t_n \right\}_{n=1,\ldots,N}$, in which each $t_n$ is a random
variable sorted  according to the density \equ{akin-equ}.
In order to get results comparable
to those of the transit times sequences, we will only allow values of
$\chi$ in the range $2<\chi<3$. For the purposes  of our computer
calculations, we generate a synthetic sequence by sampling $N$ values
$t_n$ according to the rule 
\begin{equation}
t_n=\left\{ \parbox{2in}{%
		\makebox[1.3in][l]{$t_0 \frac{\eta}{\eta_0}$} if $\eta \leq 
			\eta_0$ \\ 
		\makebox[1.3in][l]{$t_0 \left[\frac{1-\eta}{1-\eta_0}
			\right]^{-1/(\chi-1)}$} if $\eta > \eta_0$} \right. .
\label{simul-sample}
\end{equation}
where $\eta$ is a uniform random number in $]0,1]$ and
$\eta_0=1-1/\chi$. (See Appendix for details.)
Given that each term $t_n$ of any particular realization of
the sample depends linearly 
on $t_0$, we infer that the multifractal spectrum of the sequence will
be independent of the particular cut-off $t_0$ chosen. We will report
results on $D_N(q)$, the multifractal spectrum computed for an
ensemble of sequences of fixed length $N$.

When computing the  spectrum for any given value of $\chi
\in ]2, 3[$, we find that for any fixed $N$, the results for different
samples of the sequence do not collapse onto the same function, but
are widely scattered around some average position.  We explain
that effect by the fact that, by construction,  the signal
$t_n$ has no upper bound, so that it is possible to find that, just
perchance, we have generated a sample with a particular term $t_p$
extremely large, in comparison with the expected average maximum value 
$\left< T_M \right>$ (that is, a {\em rare event}). Is easy to show
(see Appendix) that in a sequence of $N$ random variables distributed
according to a truncated power-law,
the average maximum value expected scales in the limit of
large $N$ as	 
\begin{equation} 
\left< T_M \right> \sim t_0 N^{1/(\chi-1)}.
\label{cutoff}
\end{equation}
In order to get rid of the effect of those rare events, we proceed to
compute the spectrum of a {\em depleted} sequence, in which all the
values $t_n$ larger than a threshold   $\overline{T}_M \equiv t_0
N^{1/(\chi-1)}$ have been truncated 
to the value 
$\overline{T}_M$.  By using this trick, we obtain stable results
for all  sequence lengths, collapsing onto the same average curve,
within the  
error bars. 
In order to check that our particular selection of the threshold does
not have an exceedingly strong effect on the computed spectra, we have
repeated our calculations for different values of $\overline{T}_M$,
finding always the same behavior for the generalized dimensions, even
for a 
threshold as large as $t_0 N$.
In the computations reported here, we average for each
sequence length over an ensemble of $25$ different
realizations. Linear regressions were performed on intervals of $2$
decades. Statistical error bars are all smaller than $0.01$.

First of all, we observe that for $q<0$, the dual spectra are always
ill-defined, suffering from unacceptable correlation coefficients and
therefore being meaningless. This fact  seems to be very natural,
since, as it is well-known, fixed-size algorithms render bad results
for negative $q$. However, recall that what we are actually measuring
is the {\em fixed-mass} spectrum of the naive measure defined in
Sec.~\ref{naive}, so that the fixed-size spectrum of that very measure
turns out to be well behaved for {\em negative} $q$, and ill-defined
for {\em positive} $q$, against all previous intuition. 
The reason of this fact is the following: For negative
$q$ the partition function is dominated by the sparse regions of the
measure and, for positive $q$, for the dense regions. 
The bad behavior for $q<0$ is a reflection of the presence on {\em
holes} in the support of the dual measure, the only source of boxes
with abnormally small measure. 
Going back to
the naive measure, this means that this measure is dominated by a
background of a few points with an extremely large measure
(corresponding to the holes in the dual measure), which cause the
break down of the algorithm for positive $q$.
We claim therefore that the dual measure as defined in
Sec.~\ref{non-naive} is the most appropriate to characterize extremely
non-homogeneous series, like the power-law distribution under
consideration.


\begin{figure}[t]
\epsfxsize=8truecm
\centerline{
\epsfbox[0 260 530 685]{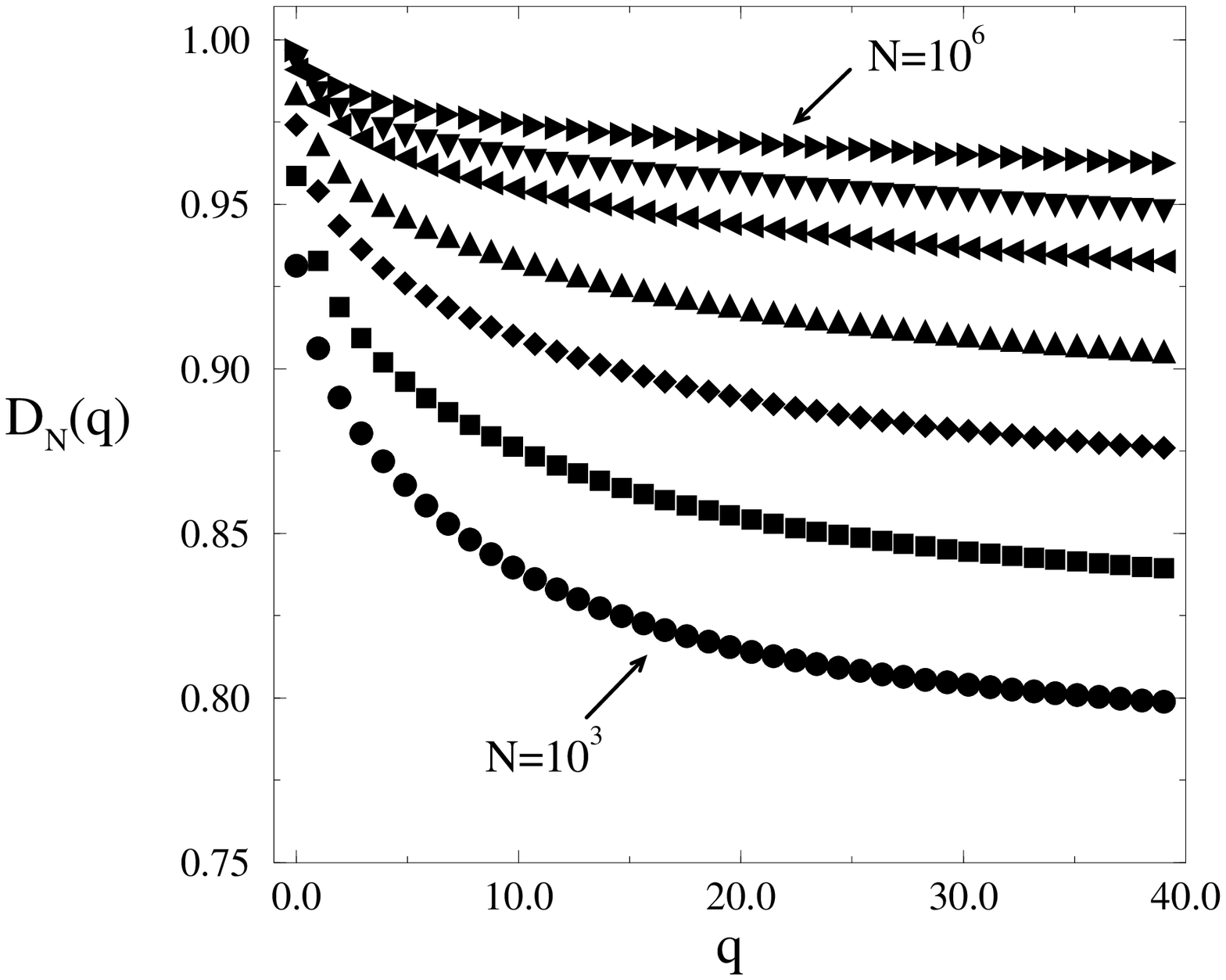}
}
\Mycap{Fig. 3:
Multifractal dual spectrum for a power-law time sequence with exponent
$\chi=2.22$. From top to bottom, the curves depict the spectrum for
sequences of length $10^6$, $3 \times 10^5$, $10^5$, $3 \times 10^4$, $10^4$,
$3 \times 10^3$, and $10^3$, respectively.
}
\end{figure}

In the range $q\geq0$, for every value of $\chi$ analyzed we observe
stable dual spectra, 
dependent on $N$, for $N>1000$. When increasing the value of $N$, the
spectrum  becomes flatter and flatter. That is to say, the
``multifractality'' of the sequence becomes  smaller and  
smaller, with $D_N(q) \to 1$ for any $q$, when $N\to\infty$. This result
is shown in Fig.~3. A measure
of the degree of multifractality ({\em
multifractality strength}), of the sequence  could
be the expression 
$1-D_N(\infty)$, where 
\begin{equation}
D_N(\infty)=\lim_{q\to+\infty} D_N(q).
\end{equation}
We have computed $D_N(\infty)$ from linear regressions of the partition
function computed for a value of $q$ large enough to ensure the
stability of the estimates. 
Numerically we find that the multifractality strength is a power-law
function of $N$,  with an exponent dependent on $\chi$
\begin{equation}
1-D_N(\infty) \sim N^{-\gamma(\chi)}.
\label{partial-scal}
\end{equation}
In Fig.~4 we have plotted $1-D_N(\infty)$ versus $N$ in log-log
scale, for different values of $\chi$. The change in the slope
is evident. In Fig.~5 we represent the estimated values of $\gamma$ as a
function of $\chi$. It is very well approximated by a linear relationship
$\gamma(\chi) \sim  \chi$.
Our numerical estimates of the coefficients of this relation are
\begin{equation}
\gamma(\chi) = (0.48\pm0.01) \chi - (0.82\pm0.02)
\label{predicted-gamma}
\end{equation}
In the limit of infinite $N$ we
will find a flat spectrum 
(uniform  measure); however, for any finite value of $N$ the
deterministic sequences are  fully multifractal.


\begin{figure}[t]
\epsfxsize=8truecm
\centerline{
\epsfbox[45 250 530 685]{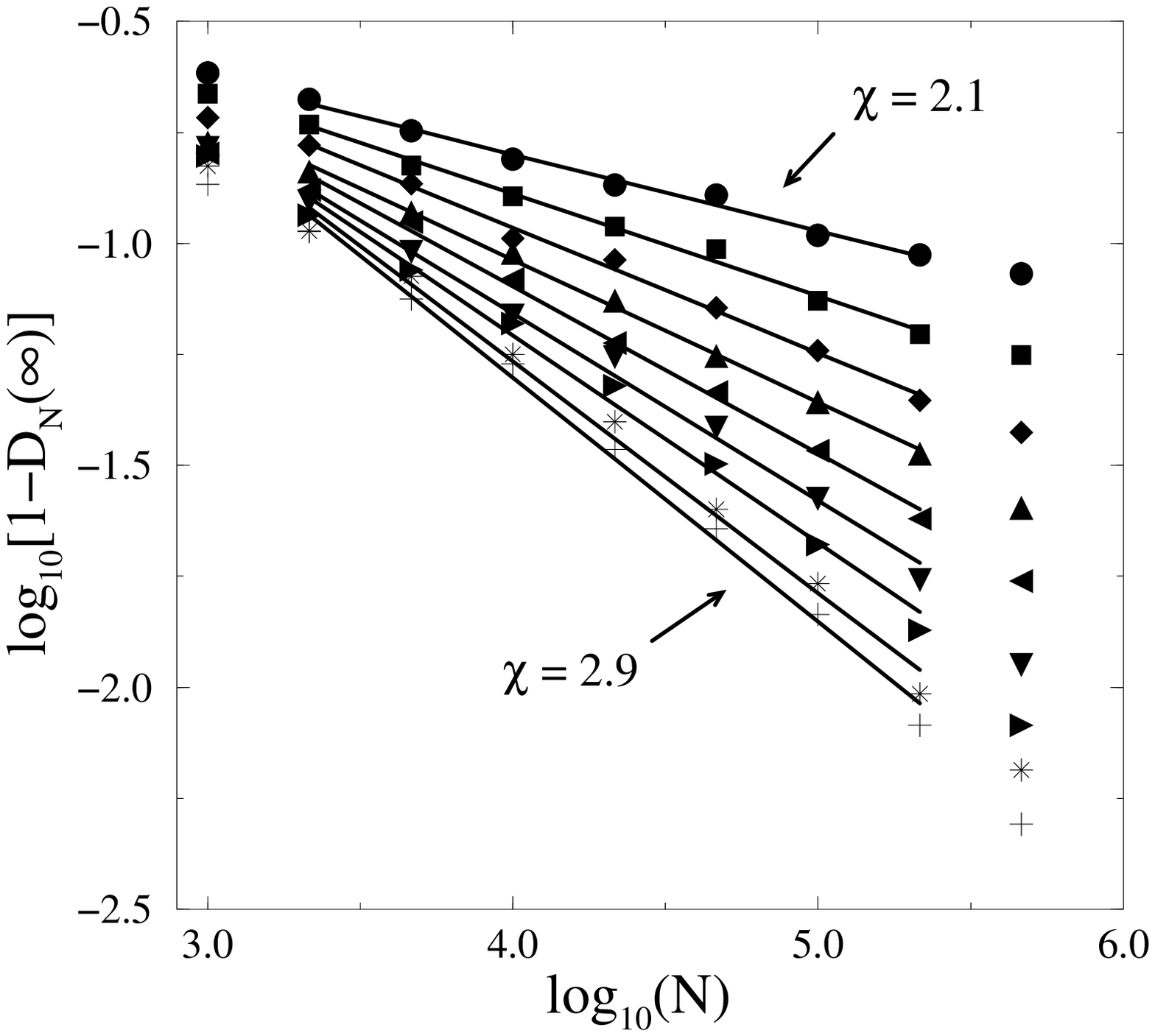}
}
\Mycap{Fig. 4:
Plot of $1-D_N(\infty)$  as a function of $N$ for $9$ values of
$\chi$; from top to bottom, $\chi$ varies from $2.1$ to $2.9$, in steps
of $0.1$. The full lines are linear fittings to the power-law
behavior.
}
\end{figure}
%


\begin{figure}[t]
\epsfxsize=8truecm
\centerline{
\epsfbox[20 260 530 685]{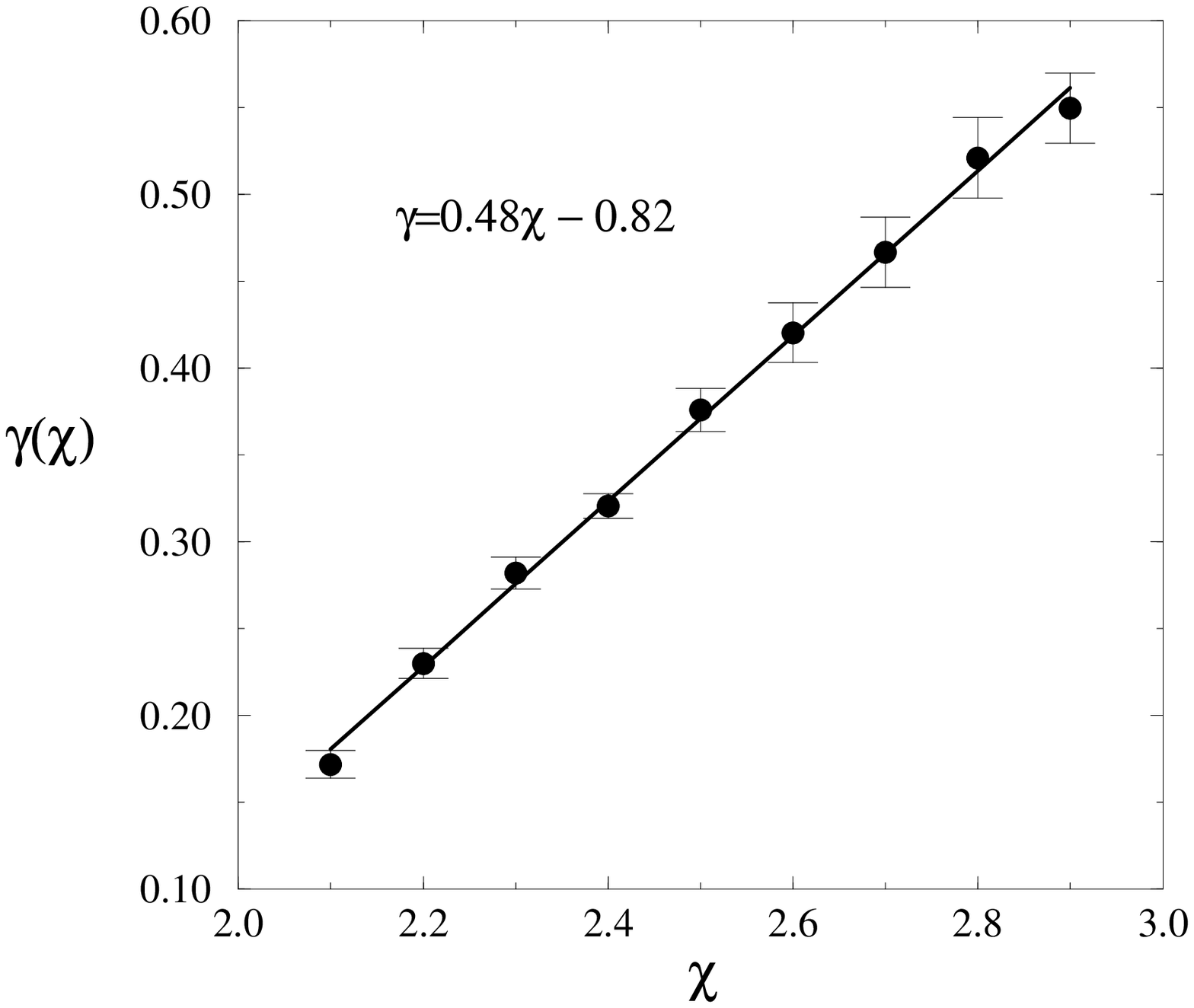}
}
\Mycap{Fig. 5:
Dependence of the multifractality strenght on the exponent $\chi$.
}
\end{figure}

Eq.\equ{partial-scal} suggests the possibility of some sort of
finite-size scaling for the multifractal spectrum: We can rewrite
\equ{partial-scal} in the form
\begin{equation}
\frac{1-D_N(\infty)}{N^{-\gamma(\chi)}} \sim \mbox{\rm const.},
\end{equation}
that is, in the limit $q\to\infty$, the spectra scales as a power law
of the sequence length. In view of this last formula, one  would be
tempted to extend the scaling to 
{\em all} values of $q$, defining a {\em renormalized} spectrum
through the expression
\begin{equation}
\frac{1-D_N(q)}{N^{-\gamma(\chi)}} = 1 - D_R(q).
\label{f-s-scaling}
\end{equation}
The renormalized spectrum $D_R(q)$ is a universal function,
independent of the length $N$. It is an intrinsic property of
the initial time sequence, independent of any particular sample, and
it can be therefore regarded as its true spectrum.

\begin{figure}[t]
\epsfxsize=8truecm
\centerline{
\epsfbox[55 260 530 685]{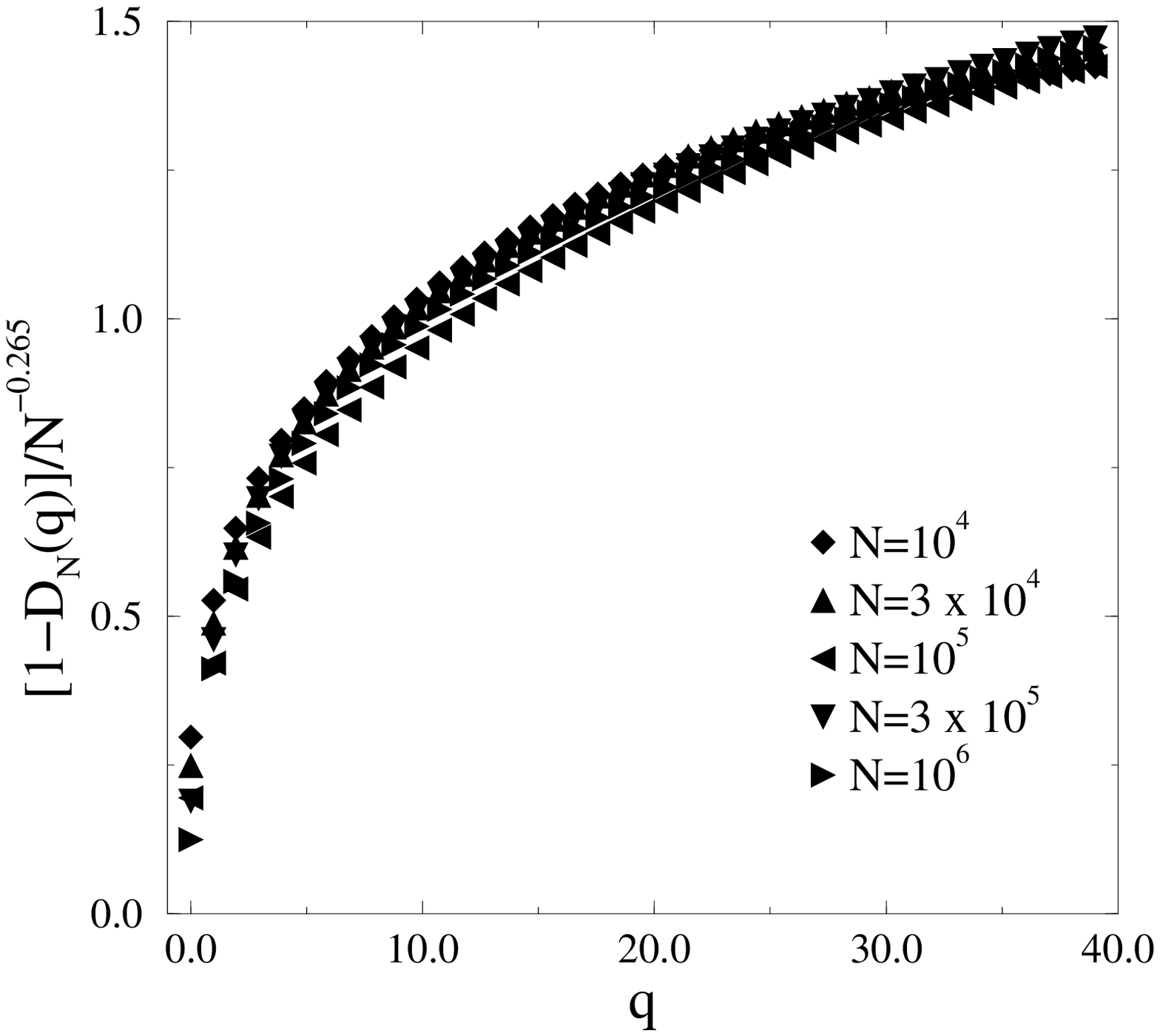}
}
\Mycap{Fig. 6:
Finite-size scaling of the multifractal dual spectrum for  power-law
time sequence with exponent $\chi=2.22$}
\end{figure}

In Fig.~6 we have tested the scaling ansatz~\equ{f-s-scaling} for
actual computations. The best collapse is achieved for sequences with
length in between $10^4$ and $10^6$, and for an exponent
$\gamma'=0.265$. The power-law sequence considered has a distribution
exponent $\chi=2.22$ and a predicted value $\gamma=0.25$ according to
Eq.~\equ{predicted-gamma}, quite close to the actual value.

\section{Numerical results for transit times sequences}

We now turn to the numerical analysis of the sequence of SOC transit
times 
$\cS$. By construction, the value $T_n$  is the
time spent into the pile  by   the $n$-th grain 
in a series of $N$ consecutive throws. 
It is conceivable that the landing of a tracer may provoke an
avalanche which would eventually evacuate out of the pile the very tracer
that caused it. In such a case, we  assign a value $T=1$ to the
transit time of that particular tracer. We have therefore $T_n \in
[1, \infty[$.
Since the computer time devoted
to any simulation is
always a finite amount, one has to stop the run
at some point, leaving inside the pile, with nonzero probability,
some of the tracers thrown at intermediate stages of the
simulation. These 
tracers which did not emerge at the end of the run would represent a
gap in the sequence $\cS$. We fill these gaps by shifting the
sequence one site to the left  at the points $n$ when a tracer did not
come out. We have also considered sequences in which each gap was
filled with a lower bound of its corresponding transit time, estimated
by substracting the time of addition of tha gap to the total time that
the simulation was running. The results obtained with both procedures
were identical, within the error bars.


\begin{figure}[t]
\epsfxsize=8truecm
\centerline{
\epsfbox[0 260 530 685]{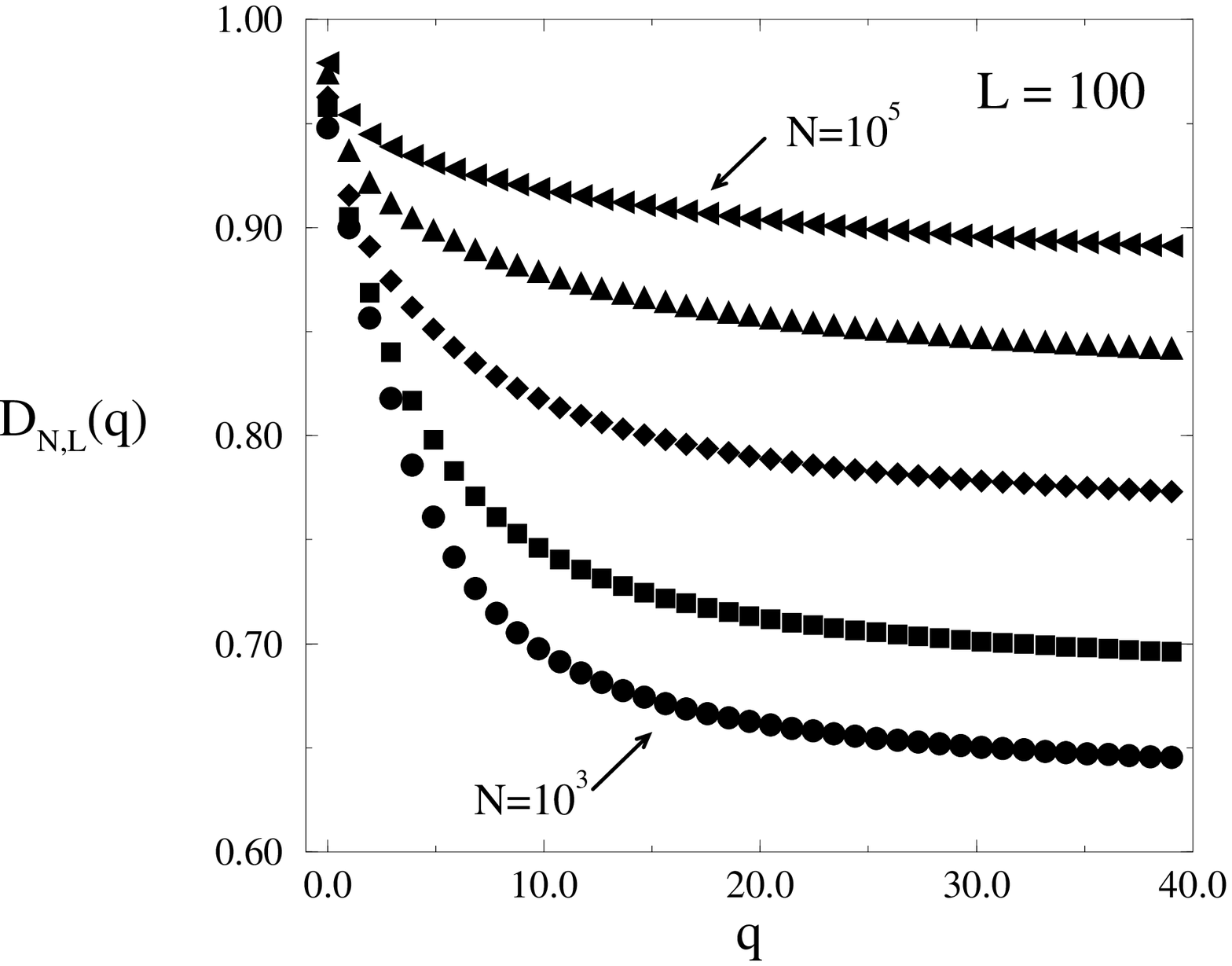}
}
\Mycap{Fig. 7:
Multifractal dual spectrum  for  SOC sequences from a ricepile of
size $L=100$. The different plots correspond to  different sequence
lengths; from top to bottom, $N=10^5$, $3 \times 10^4$, $10^4$, $3
\times 10^3$, and
$10^3$. 
}
\end{figure}
%


\begin{figure}[t]
\epsfxsize=8truecm
\centerline{
\epsfbox[0 260 530 685]{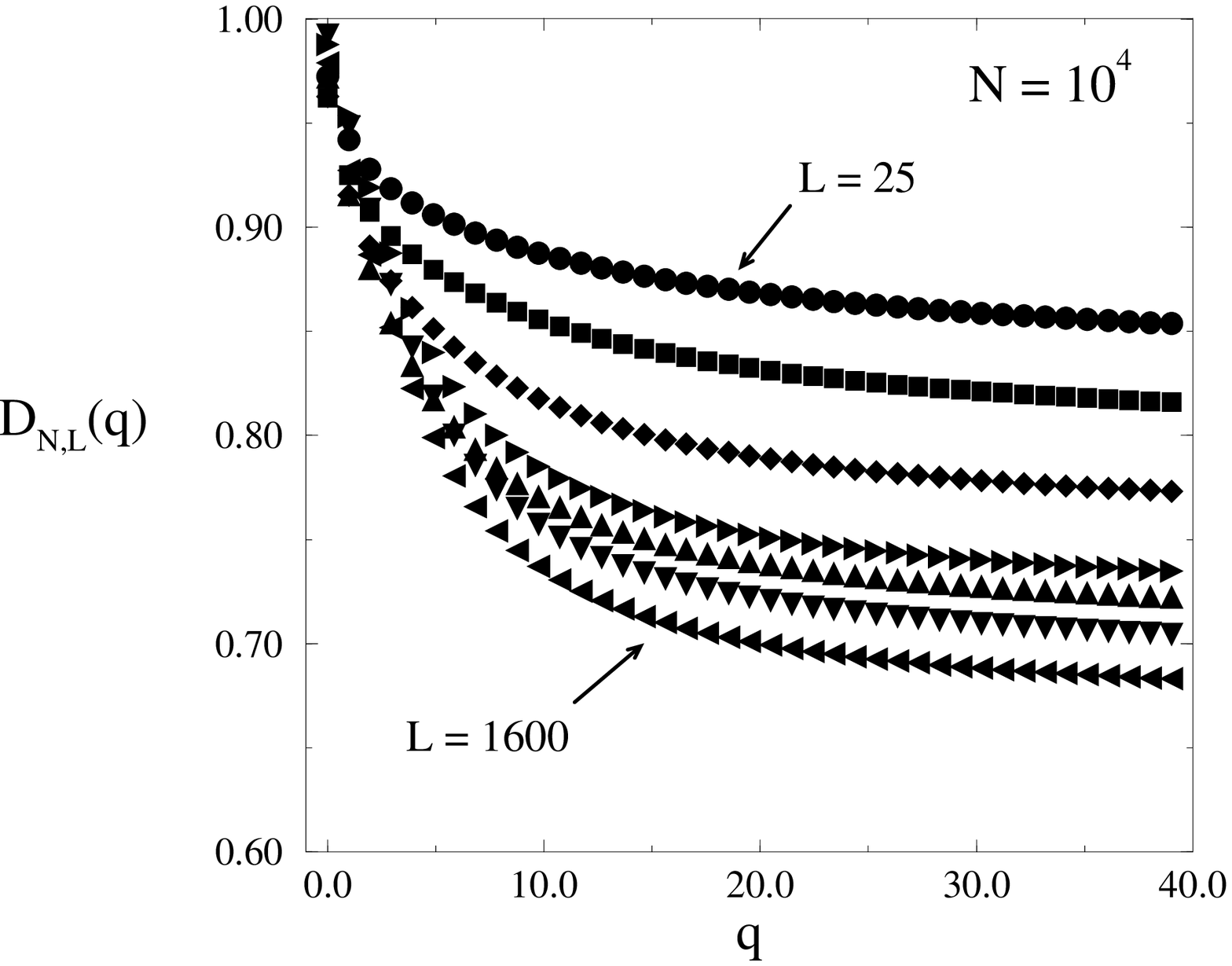}
}
\Mycap{Fig. 8:
Multifractal dual spectrum  for  SOC sequences of length
$N=10^4$. The different plots correspond to  different system
sizes; from top to bottom, $L=25$, $50$, $100$, $200$, $400$, $800$,
and $1600$.
}
\end{figure}

We work with  sequences of total length $M=10^6$ points from
simulations of the one dimensional Oslo model of size $L=25$, $50$,
$100$, $200$, $400$, $800$, and $1600$. In order to average our partition
sum, we proceed to decompose the sequences into subsequences of length
$N\ll M$, and perform the 
averages over  the sample of the resulting $M/N$ subsequences. 
When computing the spectra, however, we find that they do not stabilize
well. This is  again due to the presence of {\em rare events}: In a
subsequence of length $N$ there are some points with  extremely large
relative measure, corresponding to tracers which spent a long time
inside the pile.  In order to 
correct this effect, we proceed in the same way as we did in the random
power-law signal above: We truncate the largest events up to a
maximum 
cut-off $\overline{T}_M$. In view of Eqs.~\equ{properties}
and~\equ{akin-equ}, the SOC
signal is akin 
to a truncated power-law distributed sequence with cut-off $t_0 \sim L^\nu$;
comparing with Eq.~\equ{cutoff}, we
select $\overline{T}_M = L^\nu N^{1/(\chi-1)}$, with $\nu=1.30$ and
$\chi=2.22$, according to the simulations. 
Our results are the spectra $D_{N,L}(q)$, computed for an ensemble of
sequences of fixed length $N$, coming from a ricepile of size $L$.

With the expertise we gained from the analysis of the random power-law
signal,  we would expect the multifractal spectrum of any SOC sequence
to be ill-defined for $q<0$, to depend on the length $N$, and to be
independent of the 
cut-off, that is, of the system size $L$.
The first prediction turns out to be true; for $q<0$ the poor
correlation coefficients yield meaningless estimations. However,
for $q>0$ we obtain stable spectra depending on {\em both} $N$ and
$L$. They show an even more striking property; the spectra {\em
decrease} monotonically (become flatter) with  $N$  and {\em increase}
(become 
steeper) with $L$. This behavior is shown in Figs.~7 and~8.


\begin{figure}[t]
\epsfxsize=8truecm
\centerline{
\epsfbox[55 260 530 685]{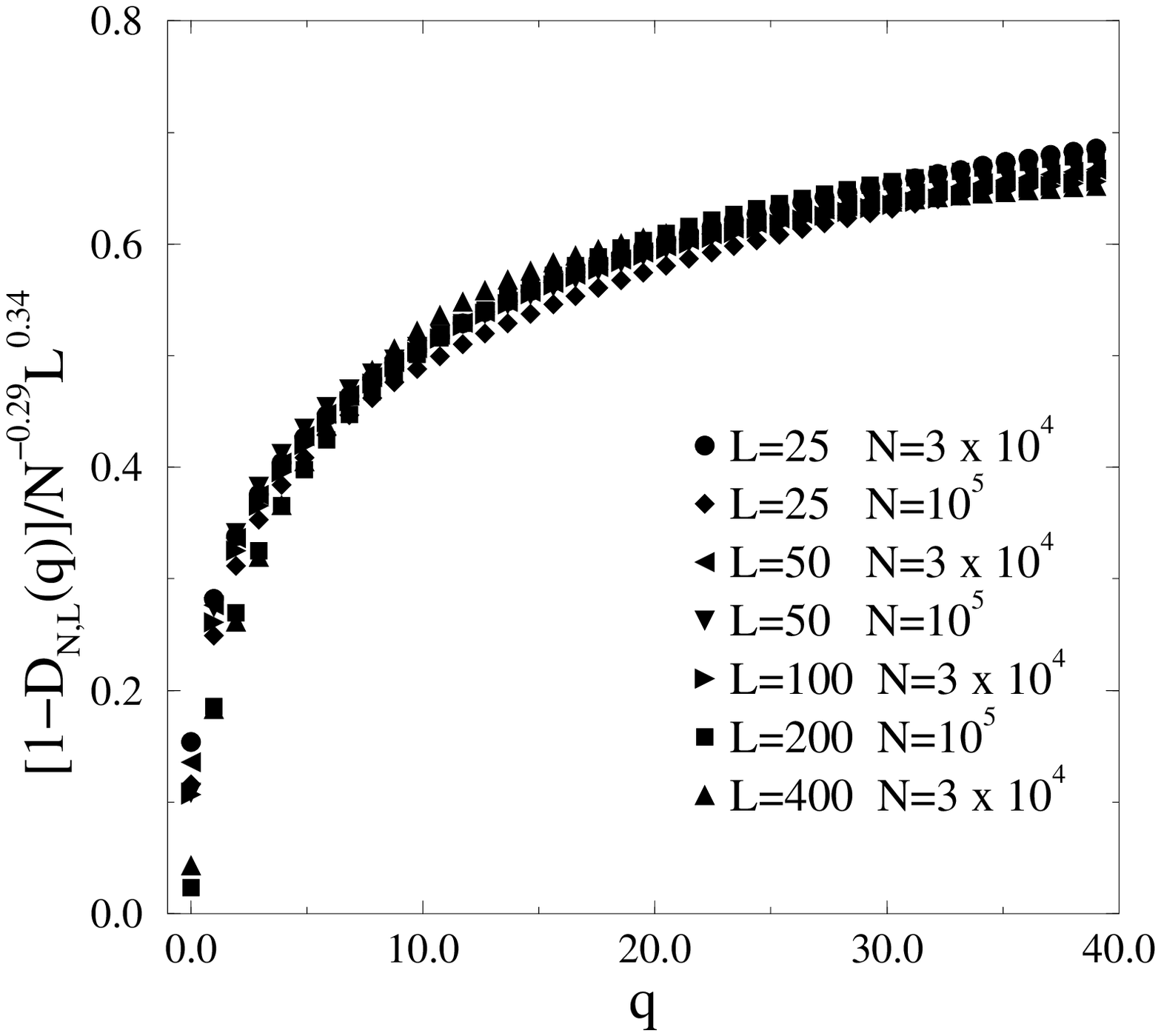}
}
\Mycap{Fig. 9:
Finite-size scaling of the multifractal dual spectrum for SOC sequences.}
\end{figure}

In a similar way as we did for the synthetic signal, we   proceed to
investigate the degree of multifractaly of the SOC
sequence. Studying the same strength parameter, we find that the
magnitude $1-D_{N,L}(\infty)$ can be fitted as a double power-law,
both in $N$ and $L$, that is,
\begin{equation}
1-D_{N,L}(\infty) \sim N^{-\gamma_1} L^{\gamma_2}.
\label{new-scaling}
\end{equation}
Our estimates are $\gamma_1=0.27\pm0.02$ and
$\gamma_2=0.32\pm0.02$. These results are valid in the range  
$N\geq10000$ and $L\leq400$.

The previous formula suggests again the possibility of constructing a
renormalized spectrum, universal for all values of $q$, and
independent of $N$ and $L$. This is done by plotting the finite-size
relationship 
\begin{equation}
\frac{1-D_{N,L}(q)}{N^{-\gamma_1} L^{\gamma_2}} = 1 - D_R(q).
\label{f-s-multiscaling}
\end{equation}
The validity of this scaling is checked up in Fig.~9. The plotted
spectra correspond to the smaller values of $L$ and larger values of
$N$ for which the relation \equ{new-scaling} holds. The best collapse
is obtained for effective exponents $\gamma'_1=0.29$ and
$\gamma'_2=0.34$, very close to the ones predicted in the limit
$q\to\infty$. 
The rescaled spectra collapse  onto a unique function, which is
interpreted again as a renormalized spectrum, in the sense that it is
a property of the intrinsic dynamics of the
ricepile where the data came from, and
independent of particular samples considered when computing it.

This  scaling behavior can be accounted for by the effect of
the correlations inside the SOC sequence. No dependence whatsoever on
the system  size (the cut-off) was observed in the synthetic power-law
distributed signal in Sec.~\ref{synthetic-power-law}. The only
difference between that signal and the SOC one resides in the {\em
correlations}. While the different points in the synthetic sequence are
completely uncorrelated by construction, the SOC transit times suffer
obviously from long-range correlations. This fact is easy to realize
when one considers that grains introduced into the pile at widely
scattered initial times can emerge at the same instant in a single
gigantic avalanche.

As a numerical experiment, we have estimated the correlation length in
our SOC sequences as the minimum length $\tilde{N}$ above which an
$R/S$ analysis \cite{feder88} provides a Hurst exponent close to
$0.5$. Our estimates show that for $L<400$ the sequences become roughly
uncorrelated for lengths larger that $\tilde{N}=10^4$, whereas no
serious estimate can be done for  $L>400$. This result seems to be in
contradiction with our multifractal scaling, since in the range of
validity of Eq.~\equ{new-scaling}, the $R/S$ analysis predicts a
complete decorrelation and, hence, an independence on the system
size. We interpret our results as a hint towards the existence of more
deep intrinsic correlations that those revealed by a simple $R/S$
analysis.

\section{Conclusions}

In this paper we have investigated the multifractal properties of
sequences of transit times of individual  grains inside an Oslo
ricepile. To this purpose, we have developed a fixed-mass multifractal
algorithm, yielding the so-called {\em dual spectrum}, particularly
well-suited to deal with highly inhomogeneous one dimensional measures
(in our case, time series). This is particularly for the
transit time sequences, which are power-law distributed and are hence
constituted at any length scale by a more or less average flat
background, interspersed by relatively infrequent huge peaks.

The main result of our analysis is the finite-size scaling 
relation~\equ{f-s-multiscaling}. 
This scaling  shows a particular behavior: The dual
spectrum tends to decrease when increasing the sequence length $N$,
whereas it tends to increase with the systems size $L$. While the
first statement is in complete agreement with  numerical
experiments on synthetic uncorrelated power-law sequences, the second
constitutes a completely unexpected result: As we show in
Sec.~\ref{synthetic-power-law}, the spectra of an uncorrelated random
power law signal do not depend on the distribution's cut-off. Since
the cut-off is related to the system size of the ricepile, we should
expect in the SOC case to obtain results independent of $L$. That is
not the case, however, in our computations. 

The renormalized spectrum defined in \equ{f-s-multiscaling} allows one
to get rid of those finite-size effects, and constitutes a magnitude
that can be associated to the very ricepile dynamics, not influenced
by the hazards of the samples used in its estimation.

We interpret the initial $L$ dependence  as an effect of the
extremely 
long correlations in the transit time sequence. As the authors point
out in Ref.~\cite{christensen96}, the fact that the average speed of
the tracers decreases with the system size proves that there are
correlations all along the system. These correlations show up even
more spectacularly  when analyzing the multifractal properties of
the sequences. A simple $R/S$ analysis seems to show an absence of
correlations for  $L<400$ and  $N>10^4$.  Hence, it could  seem
reasonable that, for these values of the parameters, the spectra
should become independent of $L$. This is not the case, however. We
conclude, therefore, that the transit time sequences indeed posess
correlations of a range far  larger than that possibly revealed by
the  $R/S$ 
analysis,  correlations which are made evident only in our more
sophisticated multifractal analysis.

\acknowledgements

I am very indebted to Alvaro Corral for providing the simulational
data analyzed in this paper and for many clarifying discussions at the
first stages of this work.
The manuscript benefited
greatly from critic readings by Jens Feder, Jordi Mach, Rudolf
Riedi, and Daniel H. Rothman.
This work has been financially supported by a scholarship grant from the
Ministerio de Educaci\'on y Cultura (Spain).

\appendix
\section*{}
In this appendix we derive some useful properties of a truncated
power-law random variable. Consider a random variable $t$ distributed
according to the density~\equ{akin-equ}. Continuity of the density at
$t=t_0$ imposes the actual form
\begin{equation}
\rho(t, t_0)=\left\{\parbox{2in}{%
			\makebox[1.1in][l]{$a\; t_0^{-1}$}
			$t\in[0,t_0[$\\
			\makebox[1.1in][l]{$a\; t_0^{-1}
			\left(\frac{t}{t_0}\right)^{-\chi}$}
			$t\in[t_0, \infty[$} \right. .
\label{appen-1}
\end{equation}
If $\chi>1$, then the density is normalizable, with a normalization
constant 
\begin{equation}
a^{-1}=\int_0^{t_0}  \frac{t}{t_0} \frac{d t}{t_0} +
\int_{t_0}^\infty \left(\frac{t}{t_0}\right)^{-\chi} \frac{d t}{t_0} =
\frac{\chi}{\chi-1}.
\end{equation}
If we demand
that the first moment of 
the distribution does exist, we have to impose $\chi>2$ to obtain
\begin{equation}
\left< t \right> = \int_0^\infty t\; \rho(t, t_0)\;dt =
\frac{1}{2 \lambda} t_0 \sim t_0,
\end{equation}
where $\lambda=(\chi-2)/(\chi-1)$.

The distribution function $P(t, t_0)=\int_0^t \rho(t, t_0) dt$ has the
form
\begin{equation}
P(t, t_0)=\left\{ \parbox{2in}{%
		\makebox[1.1in][l]{$\frac{\chi-1}{\chi}
		\frac{t}{t_0}$} $t\in[0,t_0[$\\
		\makebox[1.1in][l]{$1-\frac{1}{\chi}
		\left(\frac{t}{t_0}\right)^{-\chi+1}$} $ t\in[t_0,
		\infty[$ } \right. .
\end{equation}
In order to sample a sequence according to this distribution, we
use the {\em inversion method} \cite{bratley87}: We equate the
distribution function to a uniform random number $\eta$ and obtain the
corresponding value of $t$ by inverting $P(t, t_0)=\eta$. It is easy
to check that the resulting sample is given by Eq.~\equ{simul-sample}.

Consider now that we sort $N$ independent random variables according
to the distribution $\rho(t, t_0)$, obtaining the sample $\{t_1,\ldots
t_N\}$. Define $T_M$ as the maximum value in this particular sample,
$T_M = \max \{t_1,\ldots t_N\}$. We want to compute the average value 
$\left< T_M \right>$, weighted with the density~\equ{appen-1}.  It is
easy to see that the probability of this 
maximum value being lesser or equal than $T_M$ is just equal to the
probability of all the individual values $t_n$ being on their turn
lesser or equal than $T_M$. This means that the distribution function
of the maximum value $T_M$ is just
\begin{equation}
\Pi(T_M, N) = P(T_M, t_0)^N.
\label{appendix1}
\end{equation}
By differentiating
Eq.~\equ{appendix1} we get the probability density of maximum values

\end{multicols}

\widetext

\Raya\hfill

\renewcommand{\theequation}{A\arabic{equation}}

\begin{equation}
\pi(T_M, N) = \frac{d \Pi(T_M, N)}{d T_M} = 
\left\{ \parbox{3.8in}{%
	\makebox[2.8in][l]{$N \left(\frac{\chi-1}{\chi}\right)^N
	\left(\frac{T_M}{t_0}\right)^{N-1} t_0^{-1}$,}
		$T_M\in[0,t_0[$\\
	\makebox[2.8in][l]{$N \frac{\chi-1}{\chi}
	\left(\frac{T_M}{t_0}\right)^{-\chi} 
	\left[ 1 -
  	\frac{1}{\chi}\left(\frac{T_M}{t_0}\right)^{-\chi+1}
	\right]^{N-1} t_0^{-1}$,} $T_M\in[t_0, \infty[$} \right. .
\label{appendix2}
\end{equation}
The average maximum value that we expect to observe in $N$
samples of the initial power law distribution will then be
\begin{equation}
\left< T_M \right> =  \int_{t_0}^\infty T_M\; \pi(T_M, t_0)\;dT_M.
\end{equation}
After substituting Eq.~\equ{appendix2}, we obtain
\begin{equation}
\frac{\left< T_M \right>}{t_0} =  \frac{N}{N-1}
\left(\frac{\chi-1}{\chi}\right)^N  + 
\frac{N}{\lambda\chi}
\int_0^1 \left[ 1 - \frac{1}{\chi} \xi^{1/\lambda} \right]^{N-1} d\xi.
\end{equation}
In the limit $N\to\infty$, the only contribution in the last integral
comes from values of $\xi$ very close to $0$. We can therefore
evaluate the leading behavior for large $N$ by expanding the integrand
in Taylor series, keeping only the first order:
\begin{eqnarray}
\int_0^1 \left[ 1 - \frac{1}{\chi} \xi^{1/\lambda} \right]^{N-1} d\xi
& = &\int_0^1 \exp \left\{ (N-1) \ln \left( 1 - \frac{1}{\chi}
\xi^{1/\lambda} \right) \right\} \nonumber \\
& \simeq &\ \int_0^1 \exp \left\{ -(N-1) \frac{\xi^{1/\lambda}}{\chi}
\right \}
d\xi \simeq \lambda \left( \frac{N-1}{\chi} \right)^{-\lambda}
\Gamma(\lambda) .
\end{eqnarray}
In estimating the last integral we have extended to infinity the upper
limit, approximation allowed again in the limit of large $N$.

Collecting everything we get finally
\begin{equation}
\frac{\left< T_M \right>}{t_0} \simeq  \frac{N}{N-1} \exp \left\{ -N
\ln \frac{\chi}{\chi-1} \right\} + \chi^{\lambda-1} \Gamma(\lambda)
N (N-1)^{-\lambda}.
\end{equation}
The first term decays exponentially. Hence in the limit of large $N$,
the leading behavior is given by
\begin{equation}
\left< T_M \right> \sim t_0 \; N^{1-\lambda} = t_0 \; N^{1/(\chi-1)},
\end{equation}
up to a constant prefactor, depending only on $\chi$.

\begin{multicols}{2}

%

\end{multicols}

\end{document}